\pdfoutput=1
\documentclass[conference]{IEEEtran}
\IEEEoverridecommandlockouts
\usepackage{cite}
\usepackage{amsmath,amssymb,amsfonts}
\usepackage{graphicx}
\usepackage{textcomp}
\usepackage{xcolor}
\usepackage{booktabs}
\usepackage{url}
\graphicspath{{images/}}
\newcommand{\GapReduction}{31.9\%}

\newcommand{\OursGap}{0.380}
\newcommand{\GlobalGap}{0.559}

\begin{document}

\title{LBW-NPS: A Local Band-Weighted Noise Power Spectrum Loss for Texture-Preserving Low-Dose CT Denoising}

\author{
\IEEEauthorblockN{Fatma Elif Yal\c{c}{\i}nda\u{g}\IEEEauthorrefmark{1}\IEEEauthorrefmark{2},
Mehmet Ozan \"Unal\IEEEauthorrefmark{1}\IEEEauthorrefmark{2},
Metin Erta\c{s}\IEEEauthorrefmark{2},
\.{I}sa Y{\i}ld{\i}r{\i}m\IEEEauthorrefmark{1}\IEEEauthorrefmark{2}}
\IEEEauthorblockA{\IEEEauthorrefmark{1}Department of Electronics and Communication Engineering, Istanbul Technical University, Istanbul, Turkey}
\IEEEauthorblockA{\IEEEauthorrefmark{2}Biomedical Imaging and AI Laboratory (BIAI Lab), Istanbul Technical University, Istanbul, Turkey}
}

\maketitle

\begin{abstract}
Low-dose CT denoising should suppress noise while retaining the local texture of the reference image.
Matching a pooled spectrum can overlook regional discrepancies, whereas pixel losses alone can favor overly smooth outputs.
We propose a local band-weighted noise power spectrum loss (LBW-NPS) that compares spectral energies within overlapping Hann-tapered patches, without tissue-specific loss conditioning.
Each valid patch contributes a weighted squared log-energy difference across three frequency bands, alongside an $L_1$ reconstruction term.
We compare LBW-NPS against $L_1$-only training and five auxiliary-loss baselines using the same U-Net and three seeds on DeepLesion slices with simulated low-dose acquisitions, including patch-scale ablations and lung, soft-tissue, and bone evaluations.
On 220 test slices from 31 patients, LBW-NPS reduces a local spectral gap related to the loss, on $64\times64$-pixel evaluation windows, by \GapReduction{} relative to pooled band-weighted NPS, without a statistically significant PSNR change.
This improvement is supported by lower errors relative to the reference in average NPS frequency and gray-level co-occurrence matrix (GLCM) measures not used in the loss, evaluated in homogeneous soft tissue.
Local band-power matching therefore improves regional texture agreement without changing the denoising architecture.

\end{abstract}


\section{Introduction}
\label{sec:intro}
Reducing CT exposure increases noise and makes it harder to balance noise suppression against preservation of image detail.
Deep denoisers can improve pixel fidelity while changing the fine texture of a reconstruction.
Consequently, high peak signal-to-noise ratio (PSNR) or structural similarity (SSIM) alone does not establish perceptual or diagnostic fidelity, as recent benchmarking of low-dose CT methods emphasizes \cite{eulig2024benchmarking}.
The practical question is how to constrain texture during training while preserving the denoiser's ability to recover anatomical structure.

Several approaches extend pixel losses with texture information.
Spectrum losses have been used in cascaded ResUNet denoising \cite{liu2022spectrum}, and CT-Mamba compares radial spectra of learned noise features \cite{li2025ctmamba}.
Frequency-guided CT denoising also combines NPS constraints with intensity-focused training \cite{zhang2026frequencyguided}.
Focal Frequency Loss (FFL) supports patchwise comparisons of complex Fourier coefficients with adaptive weights \cite{jiang2021ffl}.
Other objectives combine adaptive pixel and high-frequency penalties \cite{wang2023hybrid}, match intensity distributions using Pearson divergence \cite{oh2024pearson}, or rely on perceptual features \cite{yang2018wganvgg}.

\begin{figure}[t]
\centering
\includegraphics[width=\columnwidth]{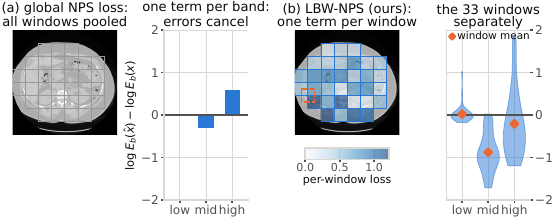}
\caption{Illustration on 33 valid windows: pooling band powers (a) can conceal regional discrepancies retained by LBW-NPS (b). Shading shows per-window loss; violins show signed log-band differences and diamonds their window means. The trained pooled comparator differs in spatial sampling and spectral estimation (Section~\ref{sec:method}).}
\label{fig:method}
\end{figure}

CT noise exhibits spatially varying magnitude and texture, depending on anatomy and reconstruction.
Studies of deep reconstruction and local NPS measurement explicitly characterize this variation \cite{solomon2020dlrecon,zhang2024localnps}.
AAPM Task Group 233 describes NPS as a measure of CT noise texture \cite{samei2019tg233}.
Anatomy-aware contrastive learning \cite{chen2023ascon}, tissue-adaptive feature modulation \cite{alfattama2026film}, and local noise-context modeling \cite{yu2026encore} incorporate related information into learned representations.
For spectral supervision, the key issue is whether regional discrepancies remain visible to the loss.
When spectra are pooled before comparison, a band-power deficit in one region can be offset by excess power elsewhere.
Agreement in the pooled spectrum can therefore coexist with local texture mismatches.

We address this issue with LBW-NPS, which compares reconstructed and reference band powers within overlapping Hann-tapered windows (Fig.~\ref{fig:method}).
Each valid window contributes a squared log-band discrepancy before averaging, so opposing regional errors cannot cancel.
Explicit band weights emphasize high-frequency mismatches, while power-only matching leaves Fourier phase unconstrained and the accompanying $L_1$ term constrains spatial correspondence.
The objective acts directly on image spectra without a learned spectral feature extractor or a change to the denoising architecture.
Reference-based body-mask and intensity criteria select valid windows; tissue identities do not condition the loss.

Our contributions are as follows: (i) a local band-energy objective for CT denoising; (ii) a comparison against $L_1$-only training and five auxiliary-loss baselines, with three patch sizes evaluated using the same U-Net architecture; and (iii) a tissue-resolved evaluation supplemented by homogeneous-region spectral and spatial texture measures.
In the evaluated setting, LBW-NPS combines the lowest local spectral gaps and high-band ratio error among the main comparisons with near-best PSNR/SSIM and second-lowest LPIPS.

\section{Method}
\label{sec:method}
LBW-NPS compares local frequency-band powers in denoised and reference images.
Let $y$ be the low-dose input, $x$ its normal-dose reference, and $\hat{x}=f_\theta(y)$ the denoised image.
For a $P\times P$ patch $z$ with mean $\bar z$, its spectrum and mean band power are
\begin{equation}
 S_z(f)=\frac{|\mathcal F\{h\odot(z-\bar z)\}(f)|^2}{\sum h^2},
 \quad E_b(z)=\frac{1}{|\mathcal B_b|}\sum_{f\in\mathcal B_b}S_z(f),
\label{eq:band-power}
\end{equation}
where $h$ is a separable nonperiodic Hann window, $\mathcal F$ the two-dimensional discrete Fourier transform, and $\mathcal B_b$ a radial frequency band.
Mean subtraction removes the constant component; Hann tapering reduces boundary leakage.
The low, mid, and high bands span 0--0.15, 0.15--0.35, and 0.35--0.5 cycles/pixel, with upper limits at 0.30, 0.70, and 1 times Nyquist.
We exclude the zero-frequency (DC) component and frequencies beyond the Nyquist radius.
Band boundaries are fixed before training and shared with the pooled baselines.
Because image spectra also contain anatomical structure, NPS matching throughout denotes this image-texture surrogate, not a noise-only spectrum.

We sample overlapping windows $\mathcal W$ with stride $P/2$; $a_p\in\{0,1\}$ indicates validity.
The default is $P=64$ with stride 32.
A reference pixel is valid if it lies in the body mask, the union of TotalSegmentator tissue masks~\cite{wasserthal2023totalsegmentator}, and satisfies $0.001<x<0.999$ in normalized intensity.
A window contributes when at least 90\% of its pixels are valid; this selection neither clips nor modifies intensities.
Batches without valid windows contribute no auxiliary loss.
Masks determine validity and crop sampling; tissue identities do not condition the loss.
Our objective averages squared log-band discrepancies over valid windows:
\begin{equation}
\mathcal L_{\mathrm{LBW}}=
\frac{\displaystyle\sum_{p\in\mathcal W}a_p\sum_{b=1}^{3}w_b
\left[\log\frac{E_b(\hat{x}_p)+\epsilon}{E_b(x_p)+\epsilon}\right]^2}
{\displaystyle\max\left(1,\sum_{p\in\mathcal W}a_p\right)} ,
\label{eq:lbw}
\end{equation}
where $\epsilon=10^{-8}$ and low-, mid-, and high-band weights $(0.5,1,2)$ are normalized to unit mean, giving high frequencies the largest weight.
Logarithms express discrepancies relative to reference energy, making bands with different powers comparable.
Squaring before averaging prevents opposite regional errors from cancelling (Fig.~\ref{fig:method}).
Power matching ignores phase; the pixel loss constrains spatial correspondence.

With one valid window covering the crop, Eq.~(\ref{eq:lbw}) becomes a single-window band loss, but still differs from pooled NPS: the latter averages selected body-ROI radial spectra before band comparison, whereas LBW-NPS averages power over two-dimensional annuli.

All models share a sampler mixing random crops with crops containing available tissue ROIs.
Training balances pixel fidelity and local spectral agreement:
\begin{equation}
\theta^\star\in\operatorname*{arg\,min}_{\theta}\;
\mathbb E_{(y,x)\sim\mathcal D}
\left[\|\hat{x}-x\|_1+\lambda\mathcal L_{\mathrm{LBW}}(\hat{x},x)\right],
\label{eq:training}
\end{equation}
where $\mathcal D$ denotes sampled training pairs and $L_1$ averages pixel errors.

The validation sweep considered $\lambda\in\{0.0007,\allowbreak 0.002,\allowbreak 0.006,\allowbreak 0.01\}$.
Our planned rule, the lowest $G_{64}$ (Section~\ref{sec:results}) within 1~dB PSNR of $L_1$, selected 0.002.
After 40 epochs, 0.002 reduced validation PSNR by 0.55~dB, with $G_{64}=0.356$; 0.0007 lost only 0.10~dB but left $G_{64}=0.417$.
We therefore departed from that rule and used 0.0007 to preserve pixel fidelity, without claiming optimality.

\begin{table*}[t]
\centering
\caption{Main results at the final checkpoint. Results are reported as means $\pm$ sample standard deviations across the three seed means. LBW-NPS combines the lowest $G_{64}$ and $G_{32}$ and the $R_H$ closest to 1 with near-best PSNR/SSIM and second-lowest LPIPS. Best values are bold; $\dagger$ marks a slice-level difference from LBW-NPS (Holm $p<0.05$ within 18 tests).}
\label{tab:main}
\setlength{\tabcolsep}{3.2pt}
\begin{tabular}{lcccccc}
\toprule
Loss added to $L_1$ & PSNR $\uparrow$ & SSIM $\uparrow$ & LPIPS $\downarrow$ & $R_H\rightarrow1$ & $G_{64}\downarrow$ & $G_{32}\downarrow$ \\
\midrule
None ($L_1$ only) & $44.35\pm0.34^{\dagger}$ & $0.9626\pm0.0011$ & $0.0585\pm0.0036$ & $0.31\pm0.04^{\dagger}$ & $0.848\pm0.011^{\dagger}$ & $1.004\pm0.033$ \\
Pooled NPS, uniform & $44.23\pm0.28^{\dagger}$ & $\mathbf{0.9627\pm0.0017}$ & $0.0476\pm0.0042$ & $0.52\pm0.08^{\dagger}$ & $0.579\pm0.007^{\dagger}$ & $0.668\pm0.007$ \\
Pooled NPS, weighted & $44.17\pm0.23$ & $0.9618\pm0.0016$ & $0.0490\pm0.0037$ & $0.52\pm0.04^{\dagger}$ & $0.559\pm0.013^{\dagger}$ & $0.653\pm0.012$ \\
FFL & $\mathbf{44.38\pm0.33}^{\dagger}$ & $0.9623\pm0.0016$ & $0.0587\pm0.0028$ & $0.31\pm0.04^{\dagger}$ & $0.849\pm0.015^{\dagger}$ & $1.006\pm0.034$ \\
VGG & $43.80\pm0.26^{\dagger}$ & $0.9476\pm0.0065$ & $\mathbf{0.0390\pm0.0031}$ & $0.78\pm0.09^{\dagger}$ & $0.551\pm0.032^{\dagger}$ & $0.613\pm0.030$ \\
Pearson & $44.14\pm0.33^{\dagger}$ & $0.9612\pm0.0018$ & $0.0586\pm0.0027$ & $0.37\pm0.06^{\dagger}$ & $0.780\pm0.033^{\dagger}$ & $0.906\pm0.042$ \\
\textbf{LBW-NPS} & $44.17\pm0.25$ & $0.9620\pm0.0021$ & $0.0418\pm0.0046$ & $\mathbf{0.84\pm0.08}$ & $\mathbf{0.380\pm0.005}$ & $\mathbf{0.490\pm0.007}$ \\
\bottomrule
\end{tabular}
\end{table*}

\begin{figure*}[t]
\centering
\includegraphics[width=\textwidth]{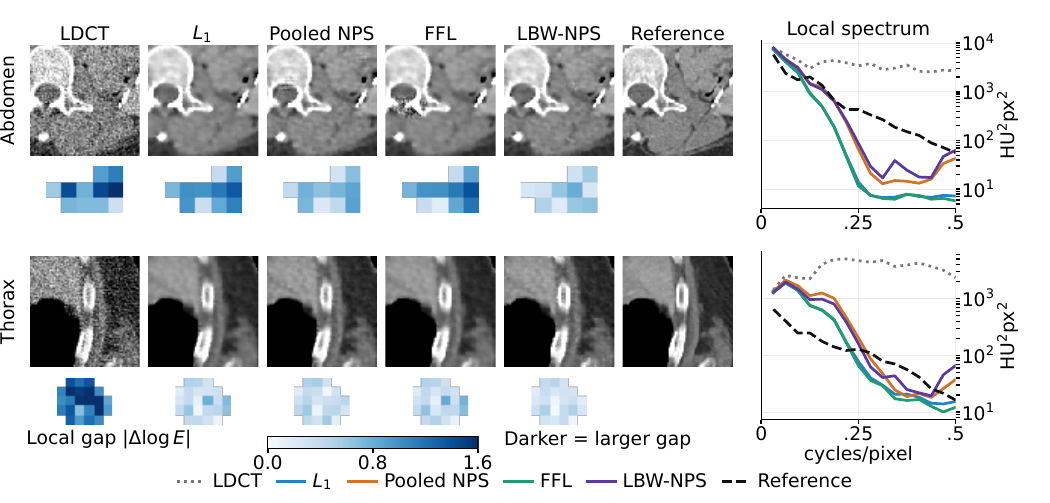}
\caption{Abdominal and thoracic examples: 96-pixel crops, local-gap maps on a shared scale, and spectra from three fixed 32-pixel soft-tissue ROIs per slice. Lighter LBW-NPS maps indicate lower local discrepancy, most clearly in the abdomen. Display window: $[-175,275]$ HU.}
\label{fig:qualitative}
\end{figure*}

\section{Experiments and Results}
\label{sec:results}
\begin{table}[t]
\centering
\caption{Scale comparison (U: uniform; W: weighted bands), reported as mean $\pm$ sample SD of the three per-seed means. $P=64$ leads on $G_{64}$; finer windows do not improve $G_{32}$ significantly at patient level.}
\label{tab:locality}
\setlength{\tabcolsep}{1.8pt}
\begin{tabular}{lccc}
\toprule
Variant & PSNR $\uparrow$ & $G_{64}\downarrow$ & $G_{32}\downarrow$ \\
\midrule
Pooled U & $\mathbf{44.23\pm0.28}$ & $0.579\pm0.007$ & $0.668\pm0.007$ \\
Pooled W & $44.17\pm0.23$ & $0.559\pm0.013$ & $0.653\pm0.012$ \\
Tissue pooling & $44.04\pm0.36$ & $0.526\pm0.029$ & $0.602\pm0.028$ \\
LBW, $P=128$ & $43.83\pm0.38$ & $0.494\pm0.045$ & $0.637\pm0.040$ \\
LBW, $P=64$ & $44.17\pm0.25$ & $\mathbf{0.380\pm0.005}$ & $0.490\pm0.007$ \\
LBW, $P=32$ & $44.22\pm0.35$ & $0.407\pm0.036$ & $\mathbf{0.488\pm0.034}$ \\
\bottomrule
\end{tabular}
\end{table}

\begin{figure}[t]
\centering
\includegraphics[width=\columnwidth]{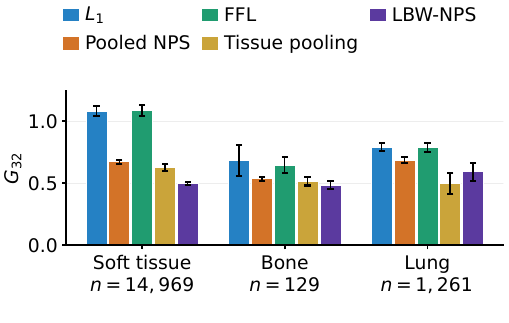}
\caption{Tissue gaps on 32-pixel windows (mean $\pm$ seed SD). LBW-NPS improves soft tissue; tissue-conditioned pooling remains better in lung. Counts denote unique spatial windows, not independent patients.}
\label{fig:tissue}
\end{figure}

Starting from the public DeepLesion dataset \cite{yan2018deeplesion}, we selected 1,301 $512\times512$ slices from 180 patients as normal-dose references.
For each slice, we simulated a low-dose acquisition by forward-projecting the reference in parallel-beam geometry, adding Poisson noise at $10^5$ incident photons, and reconstructing the noisy projections by filtered backprojection (FBP); every method denoises these FBP images toward the reference.
The patient-disjoint training/validation/test split contains 899/182/220 slices from 115/34/31 patients.
Images are normalized as $(\mathrm{HU}+1024)/4096$ over the full intensity range, without display-window clipping during training or numerical evaluation.

All models use a 7.76-million-parameter residual U-Net \cite{ronneberger2015unet}, AdamW (learning rate and weight decay $10^{-4}$), cosine scheduling, and batches of 16 $128\times128$ crops.
We train for 40 epochs of 150 iterations with seeds 42, 123, and 456, adding auxiliary losses after a 10-epoch $L_1$-only warm-up and evaluating the final checkpoint.
The six performance baselines are: $L_1$-only training, pooled NPS with uniform or weighted bands, FFL \cite{jiang2021ffl} on the full crop, VGG feature loss without adversarial training \cite{yang2018wganvgg}, and Pearson divergence \cite{oh2024pearson}.
Both pooled NPS baselines are in-house implementations, not CT-Mamba reproductions, and use $\lambda=0.002$ from our earlier experiments, without retuning.
LBW-NPS uses $\lambda=0.0007$ to preserve pixel fidelity (Section~\ref{sec:method}).
Validation-selected FFL/VGG/Pearson weights are 0.1/0.1/0.01.
Tissue-conditioned ROI pooling uses 0.002.
All three loss-window sizes $P=32,64,128$ are trained and tested at this LBW-NPS weight; $P=64$ is the main setting, and the code is available at \url{https://github.com/itu-biai/lbw-nps}.

We report PSNR and SSIM with unit data range, AlexNet LPIPS \cite{zhang2018lpips}, high-band power ratio $R_H$, and $G_q$, the mean absolute log-band discrepancy on nonoverlapping $q\times q$ valid windows with uniform band weights.
$R_H$ is the output/reference ratio of pooled radial power integrated over 0.35--0.5 cycles/pixel in matched reference-selected body ROIs, averaged geometrically over slices (ideal 1).
We evaluate body and tissue gaps at $q=32,64$; tissue scores require at least 70\% tissue occupancy.
Both body scores are reported; tissue plots and tests use $q=32$, with 129 valid bone windows versus four at $q=64$.
Paired Wilcoxon tests use seed-averaged slice metrics, with Holm correction across 18 main comparisons (six baselines; PSNR, $G_{64}$, and $|1-R_H|$).
We also test patient means to account for correlated slices.
Inference conditions on the trained models and is limited across seeds; all reported $p$-values are Holm-adjusted within their comparison family.

Table~\ref{tab:main} compares fidelity and spectral agreement across seven losses on 220 test slices.
PSNR is 44.17\,dB (0.21 below FFL; 0.18 below $L_1$), SSIM is 0.9620 (best: 0.9627), and LPIPS is second lowest (0.0418 versus VGG's 0.0390).
Against pooled band-weighted NPS, $G_{64}$ falls by \GapReduction{} (\GlobalGap{} to \OursGap{}; slice/patient Holm-adjusted $p=1.8\times10^{-33}/1.7\times10^{-8}$), without a significant PSNR change ($p=0.753/0.755$).

In Fig.~\ref{fig:qualitative}, LBW-NPS maps are lighter than LDCT in both examples and than $L_1$, FFL, and pooled NPS in the abdomen; thoracic differences among denoisers are smaller.
Both NPS losses retain more high-frequency power than $L_1$ and FFL, with frequency-dependent reference agreement.
Their similar ROI spectra place the local advantage mainly in the gap maps.

Measures not used in the loss are assessed in up to eight reference-selected homogeneous $32\times32$ soft-tissue windows per slice (197 slices, 30 patients).
Against pooled band-weighted NPS, LBW-NPS reduces average-frequency absolute error from 0.0448 to 0.0317 cycles/pixel and GLCM contrast log error from 0.696 to 0.412.
Patient-level Holm-adjusted $p$ values are $3.8\times10^{-6}$ and $1.8\times10^{-6}$, respectively, within a separate family of 27 comparisons. Against VGG, the average-frequency error does not differ ($p=0.98$), whereas GLCM contrast favors LBW-NPS ($p=3.3\times10^{-5}$).

Table~\ref{tab:locality} examines how training patch size affects spectral agreement.
$P=64$ gives the lowest $G_{64}$, whose evaluation window matches training; $P=32$ and 64 are indistinguishable on $G_{32}$ at patient level ($p=0.555$, Holm over two scale comparisons).
$P=128$ has the highest mean gaps among LBW variants, but also leaves only one window per training crop, reducing contributing windows.

To test whether the benefit extends across tissues, Fig.~\ref{fig:tissue} compares LBW-NPS with tissue-conditioned ROI pooling in soft tissue, bone, and lung. LBW-NPS improves soft tissue (0.498 versus 0.624; patient $p = 2.8\times10^{-9}$),
is worse in lung (0.593 versus 0.498; $p = 0.0012$), and shows no significant bone
difference ($p = 0.679$; Holm over three tissues).
Lung and bone each cover only 15 patients, with far fewer windows than soft tissue (Fig.~\ref{fig:tissue}).

\section{Discussion}
The results support local spectral constraints, but do not isolate pooling order as the sole cause: the pooled and local losses also differ in ROI selection and spectral averaging.
Fixed bands, patch size, and $\lambda$ remain hyperparameters; the FFL and VGG selections lie at grid boundaries, so wider searches could change their trade-offs, and FFL was not run in its patchwise mode.
Moreover, $G_q$ shares the proposed spectral construction and is not an independent noise estimate; homogeneous-region checks reduce this dependence but mainly test soft tissue.
Evaluation with a single simulated dose, dataset, and backbone leaves robustness to real acquisition noise, different reconstruction kernels, and other architectures for further investigation.
Most importantly, improved image-texture statistics do not establish lesion visibility or diagnostic benefit without task-based or reader evaluation.

\section{Conclusion}
We introduced LBW-NPS to constrain image texture through local, band-weighted spectral
comparisons during CT denoising. On simulated low-dose DeepLesion data, it reduced the
local spectral gap by \GapReduction{} relative to pooled band-weighted NPS, with no significant PSNR
difference from that comparator. In homogeneous soft tissue, LBW-NPS also agreed more
closely with the reference on two measures not used in the loss: average NPS frequency and
GLCM contrast. Of the tested patch sizes (32, 64, and 128 pixels), 32 and 64 were practical
settings for this protocol, although they do not establish an optimal patch size. In future work, we plan to separate noise from
anatomical power more explicitly, examine adaptive band weights, additional dose levels,
and other backbones, and assess whether the observed texture changes improve clinical
reading tasks.

\section*{Acknowledgment}
This retrospective study uses de-identified public DeepLesion images \cite{yan2018deeplesion}; no new participant recruitment or data collection was performed.

No funding was received for this study, and the authors have no conflicts of interest to disclose.


\end{document}